\documentstyle[eqsecnum,aps,amstex,epsfig,float]{revtex}

\begin{document}               
\draft
\title{Detection  of gravitational wave bursts by interferometric detectors}

\author{Nicolas Arnaud
\footnote{Also at Ecole Nationale des Ponts et Chauss\'ees, 6-8 av. B. Pascal, 
Cit\'e Descartes,  77455 Marne la Vall\'ee (France)}, 
Fabien Cavalier, Michel Davier and Patrice Hello}

\address{ Laboratoire de l'Acc\'el\'erateur Lin\'eaire, B.P. 34,B\^atiment 200,
Campus d'Orsay,
91898 Orsay Cedex (France)\protect\\}

\maketitle

\begin{abstract}
We study in this paper some filters for the detection of burst-like signals 
in the data 
of interferometric gravitational-wave detectors. We present
first two general (non-linear) filters with no {\it a priori} assumption on 
the waveforms to detect. A third filter, a peak correlator, is also introduced
and permits to estimate the gain, when some prior information is known about the 
waveforms. We use the catalogue of supernova gravitational-wave
signals built by Zwerger and M\"uller in order to have a benchmark of the 
performance of each filter and to compare to the performance of the optimal
filter. The three filters could be a part of an on-line triggering in interferometric 
gravitational-wave detectors, specialised in the selection of burst events.
\end{abstract}



\baselineskip = 2\baselineskip

\section{ Introduction}

Long baseline interferometric gravitational-wave detectors such as LIGO 
\cite{ligo}, VIRGO \cite{virgo}, GEO600 \cite{geo}
or TAMA300 \cite{tama} are now in their phase of construction, and should 
be fully operational in the first years of the next millenium.
Sources of gravitational waves that are expected in the bandwidth of these 
detectors, 
all involve compact objects such as black holes (BH) or neutron stars (NS);
see \cite{thorne87} or \cite{bona} for a review. Among them, inspiraling 
binaries seem to be the most promising sources for a first direct detection
of gravitational waves. Accordingly, a huge effort has been done up to now in 
order to be ready in time for analysing the inspiraling binaries 
data delivered by the interferometric detectors : detection of the signal by 
correlation with suitable templates (matched filtering), 
see e.g \cite{schutz,dhu1,apo,sathya}
or by time-frequency analysis \cite{krolak,jmi},
and estimation of astrophysical parameters (mainly the masses of the stars and 
their spins) \cite{fin,bala,dav}. 
This part of gravitational-wave data analysis, concerning
inspiraling compact binaries,  is now well advanced and rather well understood. 
Besides, periodic sources such as rotating neutron stars are maybe as
interesting as the binary inspirals, since, despite the low expected gravitational 
waves amplitudes, these sources have the advantage of being permanent.
A number of studies has been also dedicated to the analysis of periodic sources : 
see e.g \cite{schutz} or \cite{brady} and references therein.

Historically, supernovae (SN) were the first envisaged gravitational-wave 
emitters and first resonant detectors have been designed to be sensitive around 
the typical frequencies expected in such bursts of gravitational radiation, 
around 1 kHz. With the construction of intrinsically broadband inteferometric 
detectors, this kind of sources has not been studied as much as 
 inspiraling binaries or pulsars. 

Expected sources of burst gravitational waves are first collapses of massive 
stars to neutron stars (type II SN) or to black holes. 
Modern simulations of the former show a rather small 
efficiency of gravitational radiation emission 
\cite{monchmeyer,bona1,zwerger,rampp} : amplitudes typically less than
$10^{-22}$ are expected for SN at the distance of the Virgo cluster. 
Nevertheless, observation of very fast pulsars in the Galaxy (such as the 
one in the Guitar Nebula
\cite{Cordes}) may indicate that, at least in some cases,
the collapse can be highly asymmetric and provides much higher gravitational 
wave strain amplitudes \cite{Nazin}. 
Estimates of gravitational wave amplitudes from the collapse to a BH reach 
similar orders of magnitude as for previous 
type II SN  \cite{stark,bona}.

Another possible source of gravitational wave bursts occurs during the merging 
of two compact stars, at the very end of the binary inspiral. 
If the inspiral signal for binary neutrons stars
is well understood up to the 2.5 Post-Newtonian order \cite{blanchet}, we 
know only  little 
about the signal waveform corresponding to the merging phase itself, 
since its computation requires in particular fully relativistic hydrodynamical 
codes, 
although some semi-classical attempts have already been performed, 
see e.g. \cite{ooh}. 
Some recent estimates \cite{ruf} give a  maximum amplitude $h \sim$ 
a few $10^{-21}$ at 10 Mpc within a frequency range of 1-2 kHz. This is just 
the order of magnitude of the noise level at these frequencies for interferometric 
detectors in their initial design; that leaves some hope for a future detection.
Concerning BH binaries, an ambitious program called The Binary Black Hole 
Grand Challenge Alliance \cite{gca} is underway to handle the very difficult 
task of computing the waveform of the merging phase. 

Damped oscillations of excited BH's or NS's, like baby born NS's (just after the 
collapse), can also provide gravitational waves with detectable amplitudes
\cite{bona}.
The corresponding waveforms are not really bursts like, they rather have some 
coherent structure (they look like typically a damped sine). However, their
characteristic frequencies, above hundreds of Hz to tens of kHz, 
and their short damping times, make them belong to the category of signals of 
interest in this article. Note that the frequencies and damping times are 
exactly known for Kerr BH \cite{det} and the detection of gravitational 
waves emitted by such a perturbed BH could provide a direct measurement
of both its mass and its angular momentum \cite{ech}; of course, in this case, 
matched filtering, with damped sine templates, is the more suitable method.

All these gravitational wave burst signals have the following features : short 
durations from milliseconds to seconds, frequencies from $\sim$ 100 Hz to 
few kHz
and a large range of waveforms.  Filtering of such short signals in the ouput 
of interferometric detectors should therefore be as general or robust as 
possible, and designed with (almost) no {\it a priori} knowledge on the 
waveforms; this prescription of course forbids the optimal (Wiener) filtering as 
used for 
inspiraling binaries. Such general filtering methods are then necessarily 
``sub-optimal'', in the sense that they are less efficient than the optimal 
filtering.
In this article, we concentrate on the filtering of one detector's output, which 
is the first step in a detection process, the second one being the reconstruction
of the gravitational wave signal from the filtered outputs of (at least) three 
detectors. The second step has been already studied in detail \cite{san,gursel}, 
while
the first has attracted so far little attention in the litterature. Here, we study 
three filtering ideas for the detection of bursts in the data of 
interferometric gravitational-wave detectors, two of them being
very general and the third one more specific. These methods are namely a 
``counting'' method, where we count the number of bins which are larger
than some threshold in a certain window, then a method based on the 
autocorrelation function of the detector data and finally a filtering based on the
correlation of the data with a peak generic function. For each of them, we 
develop the statistical properties : link to gaussian statistics, number of false
alarms, threshold definition  ... In order to quantify the performances of such 
filters, we use as gravitational wave signals the SN catalogue from Zwerger
and M\"uller available on the web \cite{webSN}, and compute, as a 
benchmark, for each of them,
 the maximal distance of detection obtained by the three filters; as a reference, we compare
 to the maximal detection distance calculated by  optimal filtering. For this purpose, we will use, as a model for the detector noise,
the minimum of the VIRGO sensitivity, which occurs precisely in the range of frequencies of interest.

Of course, any burst filtering is unable to distinguish a non-stationary noise 
from a real gravitational wave event~: such filterings will be sensitive to transient
noise as well as to gravitational bursts. The goal of burst filtering in one 
detector is then mainly to act as a trigger and select interesting data 
streams in order to investigate
coincidences with other detectors. It will  also be useful to identify and study  
non-stationary noise in a single interferometric detector, ultimately
 providing vetos or cleaning procedures for ``known'' non-stationary 
noise sources. 

Finally, it is stressed that a general filtering approach, such as those proposed 
below,
will be sensitive to unexpected sources and therefore may provide some insight
 into new physics.

\section{Burst filtering : some ideas}

Since we know little about the expected waveforms of burst gravitational-wave 
sources a robust filtering is required. Since such a filtering is wished to
work as an on-line trigger, it should  be fast. We study three of such filters 
in the following. Throughout the paper, we assume that the detector noise 
is white, stationary and gaussian with zero mean. For the numerical 
estimates, we chose the flat (amplitude) spectral density to 
be $h_n \simeq 4\times10^{-23} /\sqrt{\mathrm Hz}$ and the sampling 
frequency $f_s \simeq 20$ kHz, so the standard deviation of the noise is
$\sigma_n = h_n \sqrt{f_s} \sim 6\times10^{-21}$; we will note the 
sampling time $t_s=1/f_s$. The value chosen for $h_n$ corresponds 
approximately to the minimum of the sensitivity curve of the 
VIRGO detector \cite{virgosens};
around this minimum, the sensitivity is rather flat, in the range $\sim$ 
[200 Hz,1kHz], which is precisely the range 
of interest for the gravitational wave bursts
we are interested in. This validates then our assumption of a white noise; 
otherwise, we can always assume that the detector output
 has been first whitened by a suitable filter \cite{cuo}.

\subsection{Bin counting (BC)}

The principle of this first filtering method is quite simple. A data stream of 
length $T$ being given (so containing $N=T\times f_s$ data), 
we count the number of data (bins)  whose value exceed a certain 
threshold, say $s\times \sigma_n$, in unit of the noise standard deviation. 
The method
is illustrated in Fig.1. It follows the prescription about no preconceived idea 
about the waveforms to detect.
In the absence of signal, the noise being gaussian, the probability that a 
data bin $x_i$ is larger than $s\times \sigma_n$ is  : 
\begin{equation}
P(|x_i| \geq s \sigma_n) = 2 \int_s^\infty {e^{-x^2 \over 2} \over \sqrt{2\pi} } dx. 
\end{equation}
It is then straightforward that $N_c$, the number of bins above threshold,
follows a binomial distribution and the probability that $N_c = n$  is
\begin{equation}
P(N_c=n) =\binom{N}{n} 
\left[ {\mathrm erfc}\left( \frac{s}{\sqrt{2}} \right)\right]^n 
\left[ 1-{\mathrm erfc}\left( \frac{s}{\sqrt{2}} \right)\right]^{N-n},
\end{equation}
where ${\mathrm erfc}$ is the complementary error function.
Setting $p = {\mathrm erfc}(s/\sqrt{2})$, the mean of $N_c$ is then 
$\mu_c = N p$ and its standard deviation is
$\sigma_c =\sqrt{N p\left( 1-p\right)}$. It is well known that the 
normalised random variable $\tilde{N}_c =(N_c -\mu_c)/\sigma_c$ 
behaves like a normalised
gaussian variable, as soon as $Np > 5$ and $N(1-p) > 5$ \cite{binom}.
These conditions will be fullfilled in every situation of interest, so we will consider
now that the random variable $\tilde{N}_c$ is well approximated by a standard normal one.

Two parameters are involved in this method : the window length $T$ 
(or equivalently $N$) and the thresold $s$.
We will discuss the window length later, with the other filtering methods. 
On the contrary,  the choice of $s$ is a specific issue for this approach.
First, $s$ should not be too large, as we expect low amplitude signals. 
Then, $s$ should not be too small because then the filter would become
very sensitive to the noise fluctuations with the drawback of a huge 
number of false alarms; 
as an example, if $s=2$, $P(|x_i| \geq 2 \sigma_n)\simeq 4.6$\%, 
 giving in average  46 'counts' in a window of $N=1000$ sampled data.
The optimal value for $s$ is evaluated as follows : we compute the 
average distance of detection for the supernovae signals of the Zwerger 
and M\"uller
catalogue as a function of $s$; the calculation of this distance is explained with full details
in section \ref{section detection}. Of course, many realisations of the noise are 
generated and  the results are averaged in order
to reduce the influence of noise fluctutations.
The result can be seen in Fig.2. We choose accordingly $s\simeq 1.7$
 in all the following, but any value of $s$ in the range [1.4,2.0] would 
also be reasonnable
(giving a loss up to 10\% 
in the average distance of detection with respect to the maximum).

\subsection{Norm filter (NF)}

This method has been initiated by the remark that  white noise samplings are 
uncorrelated, while this is in general not true for a gravitational-wave signal.
So the autocorrelation $A_x(\tau) = \int x(t)x(t+\tau) dt$ of the 
detector output $x(t)$ should reveal the presence of a 
correlated signal hidden in a uncorrelated noise. However, two 
problems arise : the  computing time (depending on the window length) and 
the choice of the detection criterion. 
The norm of the autocorrelation function seems a good one, since it should
probably distinguish between  noise and noise plus signal. But the squared norm 
of the aucorrelation of a gaussian discrete variable is 
generally not a gaussian variable itself : it becomes gaussian only for very long 
data windows, as we have checked, and so it 
requires prohibitive calculation times.
We could live with non-gaussian statistics however, but we first require 
simplicity for these preliminary studies. This is why we turn our attention 
to the maximum of the autocorrelation, which is nothing but the squared 
norm of the output $x(t)$, since the autocorrelation of any function
is maximal at { zero}. For the $N$ sampled data in a window of size $T=N/f_s$, 
the output of this filter is then
\begin{equation}
A = \sum_{i=1}^{i=N} x(i)^2,
\end{equation}
where $x(i)$ is the i$^{\mathrm th}$ data in the window.
When no signal is present, $x(i)$ is pure noise and, under our assumption 
of gaussian noise, $x(i)^2$, being the square of a gaussian random variable, is
a Chi-square random variable with one degree of freedom. $A$ is then the 
sum of $N$ such Chi-square random variables with one degree of freedom, which
is a Chi-square random random variable with $N$ degrees of freedom. Its 
mean is $\mu_A = N$ and its standard deviation is $\sigma_A=\sqrt{2N}$.
If $N>30$, the random variable (related to the norm of the windowed 
detector output)
\begin{equation}
\tilde{A} = \sqrt{2 A} - \sqrt{2N-1}
\end{equation}
is very well approximated by a standard normal variable \cite{binom}. 
This fits the simplicity requirement and we will then use the output of $\tilde{A}$
as a filter. The only parameter for this filter is the window size $N$; it will be 
discussed in the next section.

\subsection{Correlation with single pulses (PC)}

Since many of the expected waveforms present one or several peaks, it 
seems judicious to use single pulses as filters. These pulses are modelised 
with truncated gaussian functions like :
\begin{equation}
f_\tau(t) =\exp\left(-{t^2 \over 2 \tau^2}\right),
\end{equation}
with $t$ lying in the range $[-3\tau,+3\tau]$, so that the function is 
truncated at about 1\% of its maximum value.
The only parameter for this set of pulse filters is the width $\tau$.
The discrete correlation between the data $x(i), i=1,...,N$ and the pulse 
can be written as :
\begin{equation}
P(N,k)= \sum_{i=1}^N x(i+k) f_\tau([i-N/2]t_s)
\end{equation}
In the absence of a signal, the output of the filter $P(N,k)$ is a 
gaussian random variable, as a sum of gaussian random variables weighted by
the pulse fonction. The standard deviation of $P$ is simply the 
square root of the sum of the squared weigths :
\begin{equation}
\sigma_P = \sqrt{ \sum_i f_\tau([i-N/2]t_s)^2},
\end{equation}
which can be recast as
\begin{equation}
\sigma_P^2 \simeq {1\over t_s} \int_{-Nt_s}^{Nt_s}  
f_\tau(t)^2 dt \simeq \sqrt{\pi} {\tau \over t_s}.
\end{equation}
We can then define a signal to noise ratio (SNR) as the filter output 
normalised by the standard deviation $\sigma_P$ : $\tilde{P}=P/\sigma_P$, where
$P$ is the maximum of the function defined in Eq.(2.6).

\subsection{Practical implementation}

The two first methods BC and NF are very easy to implement in 
practice, as we can write simple recurrence relations 
for the calculation of the filter outputs in a given
window, as function of the filter outputs in the previous window.
For instance, for the BC, the output of the filter in a window of length 
$N$ starting at the $m^{\mathrm th}$ datum $x(m)$ and ending
at the datum $x(m+N-1)$ is $N_c(m)$. The next
filter output $N_c(m+1)$ is obtained by moving the window by one bin, 
namely starting now at the datum $x(m+1)$ and ending at the datum $x(m+N)$.
The relation between $N_c(m)$ and $N_c(m+1)$ can be cast as : 
\begin{equation}
N_c(m+1) = N_c(m)+\Theta(m+1)-\Theta(m), 
\label{recbin}
\end{equation}
where $\Theta(i)=1$ if the datum
$x(i)$ is above threshold $s\times \sigma_n$ and $\Theta(i)=0$ if 
$x(i)$ is below.

Similarily, for the NF, in the window of length $N$ starting at the 
datum $x(m+1)$, the norm of the 
data is $A(m+1) = \sum_{i=m+1}^{m+N} x(i)^2$
and is simply related to $A(m)$ by $A(m+1)=A(m)+x(m+N)^2-x(m)^2$. 
This recurrence relation between $A(m+1)$ and $A(m)$ allows a very fast
calculation of the filter output $\tilde{A}(m+1)$ to be performed from the 
calculation of the previous filter output $\tilde{A}(m)$.

One advantage of this practical simplicity  for both methods allows the 
computation of filter outputs with a window moving
from bin to bin, which is not always possible (and anyway not necessary) 
in case of correlations with a predefined lattice of filters.

~

Concerning the PC, we have first to built the lattice of filter, 
depending on only  one parameter, 
the gaussian peak standard deviation $\tau$. The parameter space 
is 
the interval $[\tau_{\mathrm min},\tau_{\mathrm max}]$. 
The distance between two successive filters of the lattice is noted $\Delta\tau$ 
and the problem is to estimate $\Delta\tau$, which is {\it a priori} 
a function of $\tau$. The output of the correlation between a gaussian 
peak filter $f_\tau$ and a ``signal'' $g$ is~:
\begin{equation}
<f_\tau,g> = K \,\,{\mathrm Max }_{t'}\,\,{\int f_\tau(t+t') g(t) dt 
\over \sqrt{\int f_\tau^2(t)dt}},
\end{equation}
where $K$ is a constant.
If $g$ is itself a filter of the kind $f_\tau'$, it is easy to show that the maximal 
correlation is obtained for $t'=0$. Following \cite{san91}, 
we chose $\Delta\tau$ such that 
\begin{equation}
{<f_\tau,f_\tau> - <f_{\tau+\Delta\tau},f_\tau> \over 
<f_\tau,f_\tau>} \leq \epsilon,
\label{ratio}
\end{equation}
where $\epsilon$ is the allowed loss in the signal to noise ratio. 
Expanding the ratio of Eq.\ref{ratio} to second order in $\Delta\tau/\tau$
leads to the simple inequality
\begin{equation}
\left(\Delta\tau \over \tau\right)^2 \leq 4 \epsilon.
\end{equation}
If the gaussian  filter $f_\tau$ of width $\tau$ belongs to the lattice, then 
the next one $f_{\tau+\Delta\tau}$ is then built from
\begin{equation}
\Delta\tau = 2 \tau \sqrt{\epsilon}.
\end{equation}
Starting from the first filter of width $\tau_{\mathrm min}$, it is then 
easy to built the k$^{\mathrm th}$ filter : its width 
is $\tau_k = (1+2\sqrt{\epsilon})^{k-1} \tau_{\mathrm min}$. The total 
number of templates in the lattice 
is finally the maximal integer $n_t$ such that
$(1+2\sqrt{\epsilon})^{n_t-1} \tau_{\mathrm min} \leq \tau_{\mathrm max}$.

For example, Fig.3 shows the distribution of the 117 templates in the 
interval [1\,ms,10\,ms] for a loss in SNR $\epsilon=10^{-4}$ (the one 
we will use
in the next section); the choice of $\epsilon=10^{-2}$ reduces the number 
of filters to 13 for the same interval. We notice that
we can allow for a very low loss in SNR and still obtain a reasonable 
number of templates; this is due both to the fact that 
we deal with an one-dimensional lattice space
and to the 'smooth' dependence of $n_t$ on $\epsilon$. For instance, 
the same very low value of $\epsilon =10^{-4}$ and a parameter 
space extended to
the (physically possible) interval [1\,ms,1\,s] lead to a lattice of 
only 349 templates, which is easy to implement for an on-line processing.

\subsection{Threshold and false alarms}

Since the outputs of the three proposed filters behave like standard normal 
random variables (when no signal is present), it is convenient
to define the same detection threshold for all, with, consequently, the same 
number of false alarms produced. 
As we expect weak signals, this threshold has to be low.
On the other hand, we can deal with a large number of false alarms; these 
spurious events can be processed and discarded later when working in coincidence
with other detectors. The relation between the detection threshold $\eta$ and 
the rate of false alarms $r_{\mathrm fa}$ for a gaussian random variable~:
\begin{equation}
2 \times{1 \over \sqrt{2\pi}} \int_{\eta}^{\infty} 
\exp\left( {-x^2 \over 2}\right) dx = r_{\mathrm fa}.
\end{equation}
A false alarm rate $r_{\mathrm fa}=5\times10^{-7}$ ($\simeq 36$ false 
alarms per hour for a sampling rate $f_s = 20$ kHz) corresponds to a threshold
$\eta \simeq 4.75$, while a false alarm rate 10 times smaller corresponds to 
$\eta \simeq 5.20$. For the results presented in the following), the chosen 
threshold  is $\eta \simeq 4.75$.

For the two first filtering methods, BC and NF, the situation is however 
not so simple because the outputs of the filters in two successive windows
are in fact strongly correlated. For example, for the BC  filtering, the 
filter outputs in two successive windows (starting respectively at the datum
$x(m)$ and $x(m+1)$ are related by  Eq.\ref{recbin}; it is clear that 
$N_c(m)$ and $N_c(m+1)$ are the same or differ at 
most by $\pm 1$. So if the detection
threshold is exceeded in a number of consecutive windows, there is in 
general only one ``event''. 
This leads to redefine what is a detected event : an event is said to be 
detected in some time  interval $[m_1 t_s,m_2 t_s]$ 
if the filter output $O(m)$ exceeds the threshold $\eta$ for all values 
of $m$ in the interval $[m_1,m_2]$ and is less than $\eta$ outside. This is equivalent
to define a 'correlation length' $(m_2-m_1) t_s$ for the event to be detected.

\section{Detection of supernovae}
\label{section detection}

In this section, we use the catalogue of simulated gravitational-wave signals 
emitted during supernovae collapses 
and computed by Zwerger and M\"uller \cite{zwerger}, 
to implement and test the three filterings described previously, in a realistic context.

\subsection{The catalogue}

The catalogue of Zwerger and M\"uller \cite{webSN} contains 78 
gravitational-wave signals generated by axisymmetric core collapses. Note that, in particular due to axisymmetry, these are purely linearly polarized
waves ($h_\times =0$). Each of them corresponds to a particular 
set of parameters, essentially the initial distribution of angular momentum 
and the rotational energy of the star core,
in the collapse models of Zwerger and M\"uller. The  signal total durations 
range from about 40 ms to a little more than 200 ms. 
The gravitational wave amplitudes of the stronger signals are of the order 
$h (=h_+)\sim$ a few $10^{-23}$ for a source located at 10\,Mpc; that leaves little hope
to detect such events with the first generation detectors.
Concerning the shape of the waveforms (see Fig.4), 
Zwerger and M\"uller distinguish three different types of signals\cite{zwerger}. 
Type I signals typically present 
a first peak (associated to the bounce)followed
by a ringdown. Type II signals show a few (2-3) decreasing peaks, 
with a time lag between the first two  of at least 10\,ms. Type III signals exhibit
no strong peak but fast ($\sim$ 1 kHz) oscillations after the bounce. 
The fact that type I and type II signals are characterised by
strong peaks validates the choice of the 
filtering by correlation with generic peaks in order to detect such events.
The 78 signal templates in the catalogue are not equally sampled, so we 
have first re-sampled them by interpolation at the desired sampling frequency
($f_s = 20$\,kHz in our examples).

\subsection{Optimal filtering and maximal distances of detection}

Since the 78 signal waveforms are known, we can explicitely derive the 
optimal SNR provided by the Wiener filter for each of them, and 
then compute the
maximal distance of detection. We will then be able to build a benchmark 
for the different filters by comparing their results (detection distances) to the results
of the Wiener filter. In all the following, we assume that the incoming waves have an optimal
incidence with respect to the interferometric detector.

Let's call $h(t)$ one of the 78 signals and $\tilde{h}(f)$ its Fourier transform. 
The optimal signal to noise ratio $\rho_0$ is given by 
\begin{equation}
\rho_0^2 = 2 \int {|\tilde{h}(f)|^2 \over S_h(f)} df
\end{equation}
where $S_h$ is the one-sided noise power spectral density (hence the factor of 2). 
The noise is 
assumed to be gaussian and white with a standard deviation related to the 
constant spectral density $S_h=h_n^2$ by
$\sigma_n = h_n \times \sqrt{f_s} = \sqrt{S_h \times f_s}$. The SNR 
then becomes
\begin{equation}
\rho_0^2 = {2 f_s \over \sigma_n^2} \int |\tilde{h}(f)|^2  df 
= {2 f_s \over \sigma_n^2} \int |{h}(t)|^2  dt.
\end{equation}
As previously, a supernova signal is detected by the Wiener filter 
if $\rho_0 \geq \eta$, where $\eta$ is the same detection threshold as defined above.
Fig.5 shows the maximal distance of detection for each of the 78 
signals. The mean distance, averaged over all the signals, is about 26.1\,kpc, 
which is of the order of the diameter of the Milky Way. 
A  few signals can be detected at distances beyond 50\,kpc, 
the distance of the Large Magellanic Cloud (LMC). The most energetic 
signals (number 77 and 78) can be detected at distances as high as 100-120 kpc.
It is clear that this class of signals will be detected by the first generation inteferometric 
detectors only if the supernovae occur inside our Galaxy or in the very
close neighbourhood.

\subsection{Detection by the three filters}

\subsubsection{Window sizing}

The window size $N$ for the first two filters is a compromise between the need 
for not too small number of bins, in order to garantee that the filter outputs
are well approximated by gaussian random variables, and the rather short signals 
we are looking for. The first constraint is easily fulfilled for the NF
method, where $N$ must be larger than about 30. For the BC method, we must 
have $N {\mathrm erfc}(s/\sqrt{2}) > 5$ 
and $N (1-{\mathrm erfc}(s/\sqrt{2})) > 5$; with the optimal choice of 
$s=1.7$, so that ${\mathrm erfc}(s/\sqrt{2}) \simeq 0.91$, the two inequalities
give respectively $N>6$ and $N>56$. The  constraints are not very severe and 
require that $N$ must be between 80 and 150, as it has been checked
by Monte-Carlo simulations (looking for the optimal window size giving the 
maximal mean distance of detection).

For the pulse filter PC, the window size is automatically fixed by the definition 
interval of the gaussian peak $[-3\tau,3\tau]$ for a filter of width $\tau$. The 
window size for the correlation with the filter $f_\tau$ is then $N = 6\tau f_s$ 
and depends then on $\tau$.

\subsubsection{Detection distance}

The efficiency of each filter is measured by the maximal distance of detection 
for each of the 78 gravitational wave signals of the catalogue.
This distance is obtained by averaging over many noise realisations in a 
Monte-Carlo simulation. We present the results in two ways. Fig.6 shows the
number of detected signals as a function of the distance of detection for the 
three filters. Results of Fig.6 combined with the results for the optimal filtering
(Fig.5) are reported in Fig.7 in a normalised way.
The histograms in Fig.7
show the number of signals detected as a function of the reduced distance of 
detection for the three filters; the reduced distance of detection
is simply  the distance of detection divided by the maximal
distance of detection computed with the optimal filter. The means of the 
distributions are respectively 0.22, 0.26 and 0.34 for the 
BC, NF and PC filters; these
can be seen as rough estimators of the efficiency of the filters. 
The histograms in Fig.7 give also an idea of the sensitivity of the filters to  the
waveforms of the detected signals. We note that the histograms corresponding 
to the BC and NF filters are much more concentrated than the one corresponding 
to the PC filter; this is particularly impressive for the NF filter. 
This means that the two first filters are rather robust and their 
efficiency does not crucially 
depend on the details of the gravitational wave signals. 
At the contrary, the larger dispersion of the last histogram (PC) 
indicates that the response of the PC
filter depends much more on the gravitational waveform.

The global efficiency can be measured as the mean detection distance 
averaged over the 78 signals. The results are reported in Table 1.
The efficiency of the BC and NF filters are about one third of the 
efficiency of the optimal filtering (max. efficiency), while the PC filter 
(for $\epsilon=10^{-4}$) has an efficiency about
58\% of the maximal efficiency.
\begin{center}
\begin{tabular}{|c|c|c|c|c|}
\hline Filter  & Optimal & BC & NF & PC \\
\hline Average distance (kpc) & 26.1 & 7.8 & 9.3 & 15.2\\ \hline
\end{tabular}

{Table 1 : Detection distance averaged over the 78 signals for the different filters.}

\end{center}

We note that none of the BC, NF and PC filters is efficient enough to cover 
all the Galaxy on average. Several  signals however can be 'seen' anywhere from
the Galaxy and even beyond; in particular the signals 77 and 78 can be detected 
up to the LMC by any of the three filters.

In fact, concerning the PC filter, the mean distance of detection depends on 
$\epsilon$, the allowed loss in SNR, and 
consequently on the number of filters in the lattice.
Fig.8 shows the mean distance of detection as a function of the loss in SNR. 
As $\epsilon$ decreases, the mean distance of detection becomes closer to
the maximal value ( a little larger than 15 kpc). We also notice that for high 
values of $\epsilon$ (low number of templates), 
for instance above $\epsilon =1\%$ (13 templates), the efficiency of the PC 
filter is still well larger than
the NF/BC efficiency.

\subsubsection{Computation time considerations}

Apart form the criterion of simplicity of the filters, we require also that 
the data processing with these filters has to be fast enough to be implemented 
on-line as a trigger. 

An analysis of one day of data has been performed (in order to check the 
validity of the redefinition of an 
event -see section II.E- and the number of false alarms)
on a DEC Alpha workstation, in about 14\,mn with the BC 
filtering and in about 25\,mn with the NF filtering. So there is about a 
factor 100 (resp. 57) between the data stream duration and the 
time needed to process 
them by the BC filter (resp. the NF filter). These two filters can 
then be used on-line without any problem.

The PC filter is no more problematic due to implementation in the 
Fourier space and use of FFT's. Moreover the correlations 
don't need to be computed
in successive windows. Indeed a template $f_s$ (gaussian of width $s$) 
has a certain correlation length ($\sqrt{2}s$), 
so it is possible to compute the correlation with the template $f_s$ 
every $T_s$, where $T_s$ is related to the allowed loss 
in filter efficiency. In practice,
one order of magnitude on the calculation time can be gained with 
a loss of 1-2 \% of efficiency.

\section{Conclusion}

We have studied in this paper three filtering methods with the aim 
of building on-line triggers for the selection of burst-like events in the data flow of 
interferometric gravitational-wave detectors. Such a filtering needs 
first to be as general and as simple as possible; these two prescriptions 
are well fulfilled
by the two first : BC and NF filterings. A third filter has then be studied 
in order to quantify what could be the gain with a more specific filtering, i.e. using
some {\it a priori} information on the signals to detect (namely here 
the fact that the supernova signals contain at least one peak of short duration).
The catalogue of supernova gravitational-wave signals provided by 
Zwerger and M\"uller has been used in order to built a benchmark for the different
filters in a realistic way, and to compare each of them to the optimal (Wiener) filter.

The results we find is first that a general filter, such as BC of NF, has 
an efficiency of about one third (0.30-0.36) 
of the efficiency of the optimal filter (maximal efficiency for a linear filter). 
The peak correlator PC (looking for the peaks in the Zwerger and 
M\"uller signals) has a efficiency slightly larger (58\%
of the optimal filter efficiency). We note also that the general filters 
seem to be more robust (less sensitive to the waveform details)
than the peak correlator, because of
the smaller dispersion of their (normalized) responses to the 78 supernovae signals.

Concerning a practical implementation, each of the filters can be 
implemented on line and can be used as a trigger in order 
to select events for off-line coincidences with other detectors.

The results for the detection of the Zwerger and M\"uller signals 
are not very optimistic.  
Moreover, as we have assumed optimal incidence for the gravitational waves, the detection
distances should be divided by $\sqrt{5}$ for sources randomly
distributed over the sky \cite{thorne87}. So only a small fraction of the 
supernova events can be detected
anywhere in the Galaxy or beyond. This is not a surprise, and, 
anyway, not the main point of this paper.

We finally notice that the two general burst filters BC and NF 
we studied are in fact non-linear filters (they can not be reduced to a 
correlation with the detector output). This may encourage people to 
develop and study such a class of filters in the context of gravitational 
wave detection.
In the near future, we plan to study other filters based on the autocorrelation 
function and on wavelets analysis. We plan also to study the effect of a more 
realistic noise on the response of this class of filters, in particular the effect of 
non-gaussianity and non-stationarity. 
We keep in mind that non stationary noises can be treated  as 'signals' and the 
burst filtering can help to identify them and
finally understand the detector behavior.

\baselineskip = 0.5\baselineskip  

 
\newpage
\begin{figure}
\caption{Principle of the Bin Counting filter. The filter select all bins that 
are above some level ($3\sigma_n$ on the example); here a signal, starting at time
0.5 s has been superposed to the noise.}
\end{figure}
\begin{figure}
\caption{Efficiency of the BC filter as a function of the threshold level $s$. 
The efficiency values have been normalised to the maximum value.}
\end{figure}
\begin{figure}
\caption{Distribution of the 117 templates for the peak correlator in the 
interval [1;10] ms with the assumption $\epsilon=10^{-4}$.}
\end{figure}
\begin{figure}
\caption{Typical  waveforms for the type I, II and III supernovae signals 
in the Zwerger and M\"uller catalogue.}
\end{figure}
\begin{figure}
\caption{Detection distances calculated with the optimal filter for the 78 signals 
in the catalogue of Zwerger and M\"uller. About 6 signals can be detected
at distances as high as about 50 kpc (the LMC distance).}
\end{figure}
\begin{figure}
\caption{Histogram of the number of signals as a function of the maximal detection 
distance for each of the three filters ($\bullet$ : BC, $\circ$ : NF and
$\star$ : PC).}
\end{figure}
\begin{figure}
\caption{Histogram of the number of signals as a function of the normalised 
detection distance for each of the three filters. The normalised distance is
 the detection distance divided by the corresponding maximal detection 
distance computed by the optimal filtering.}
\end{figure}
\begin{figure}
\caption{Efficiency of the Peak Correlator PC  versus the loss in SNR $\epsilon$. 
The efficiency is 
measured as the mean detection distance for the 78 signals
in the Zwerger and M\"uller catalogue.}
\end{figure}
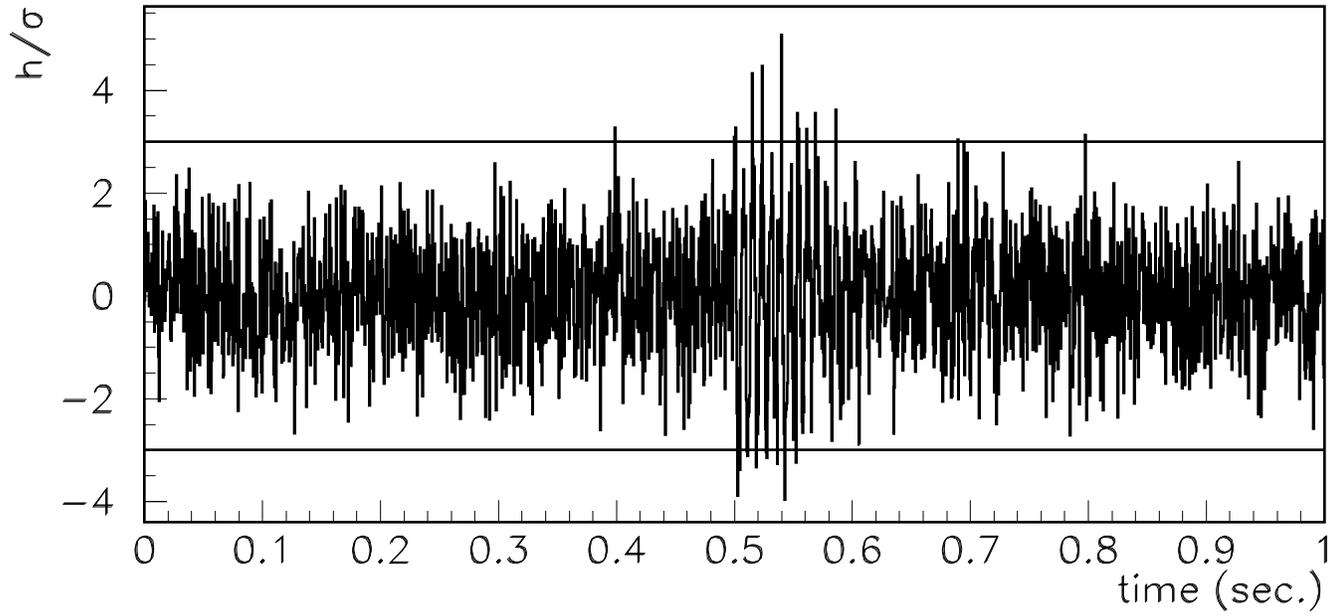
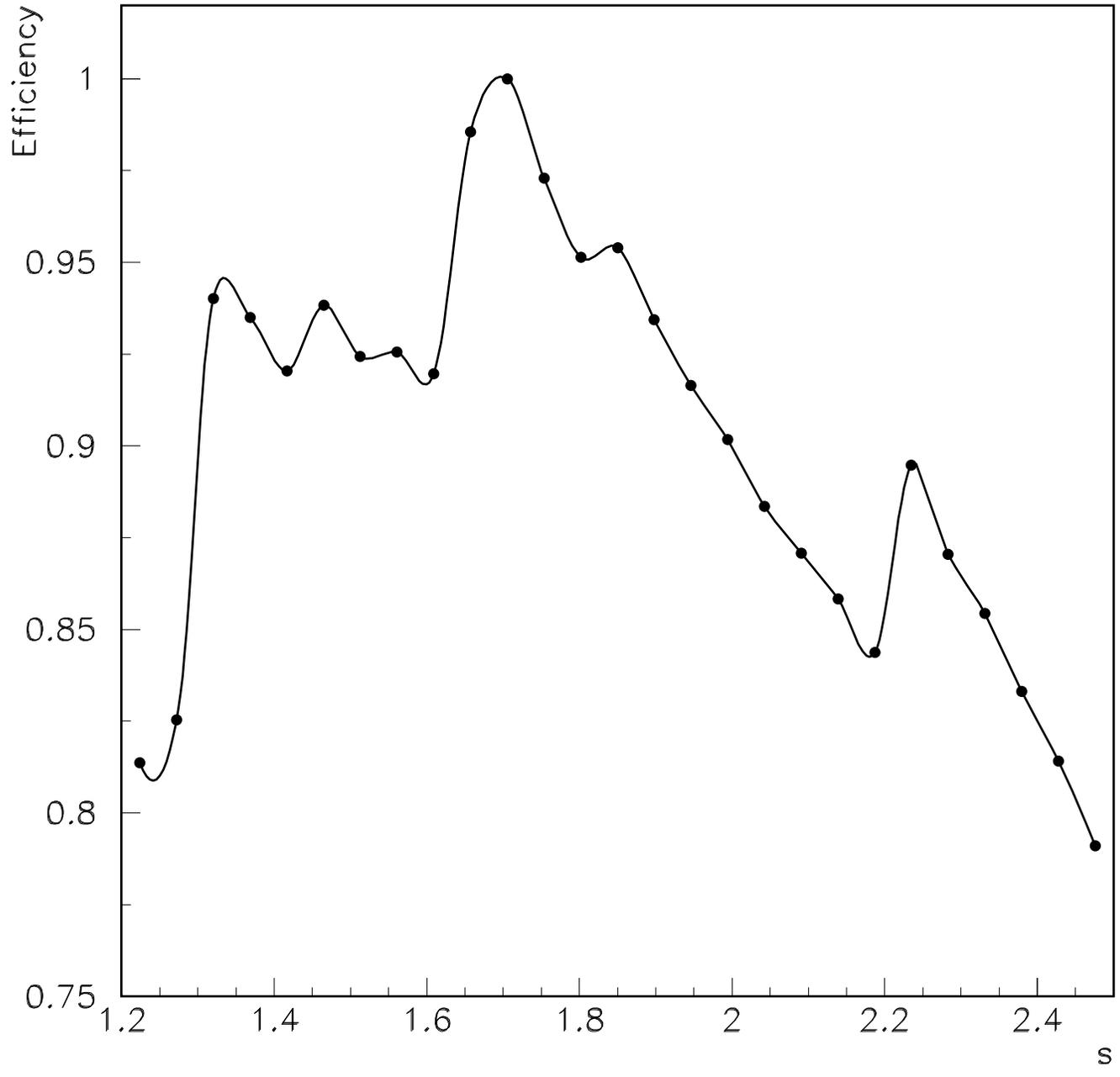
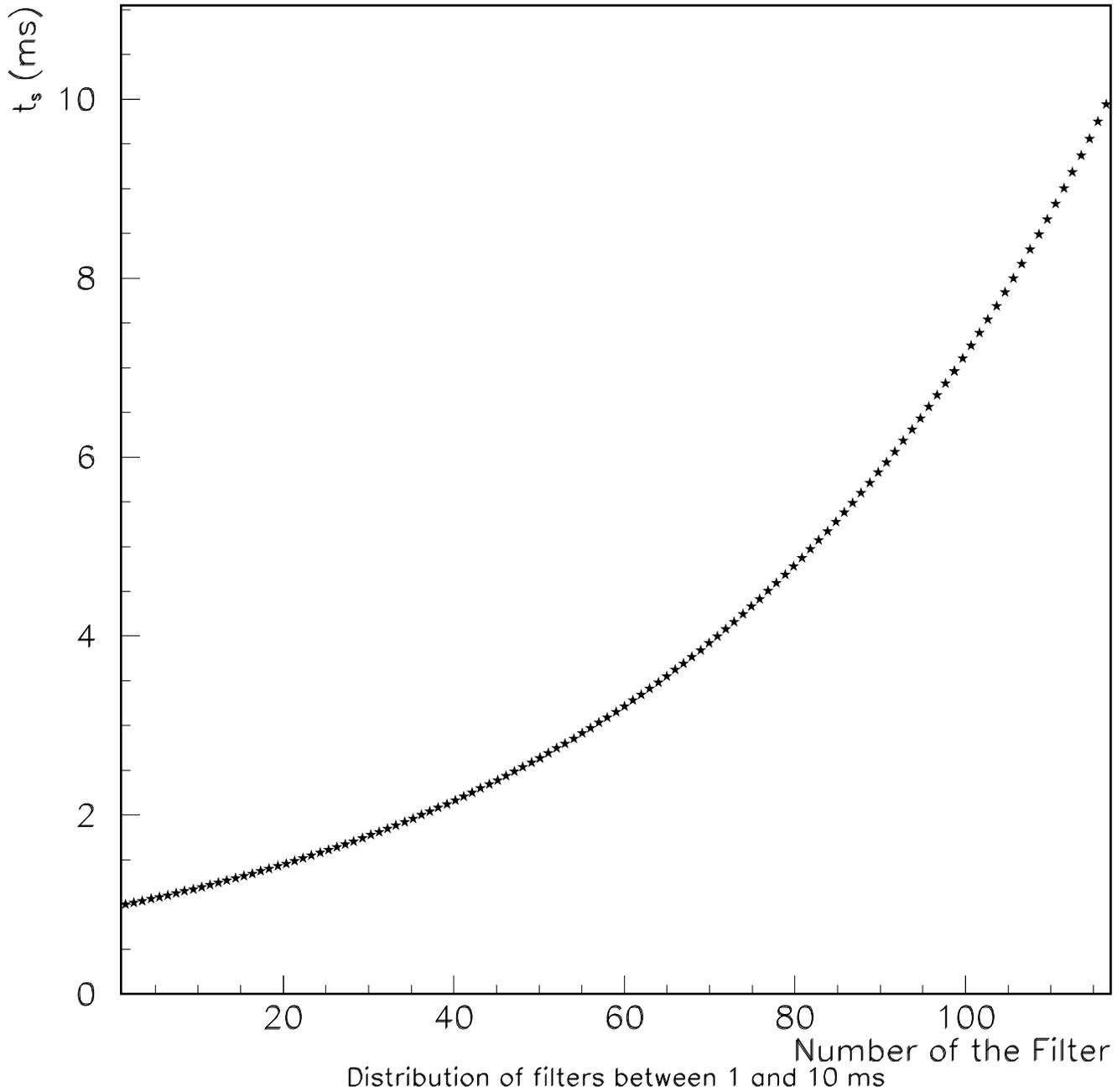
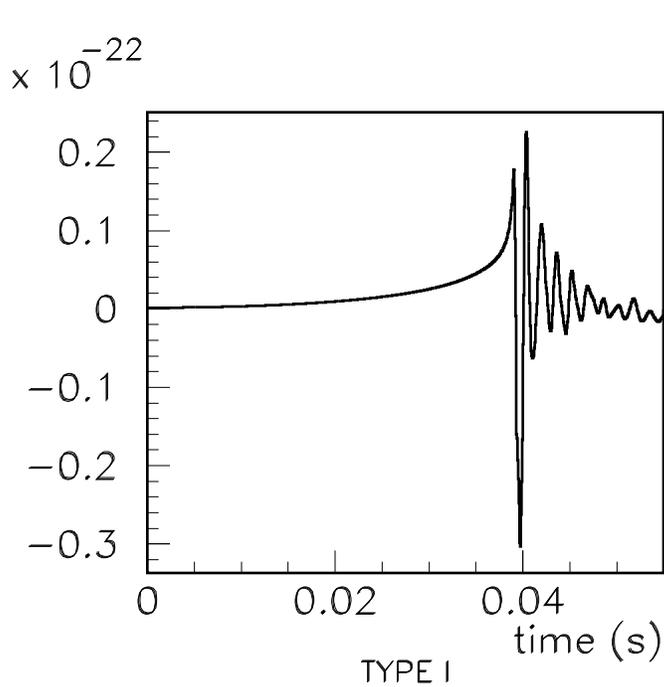
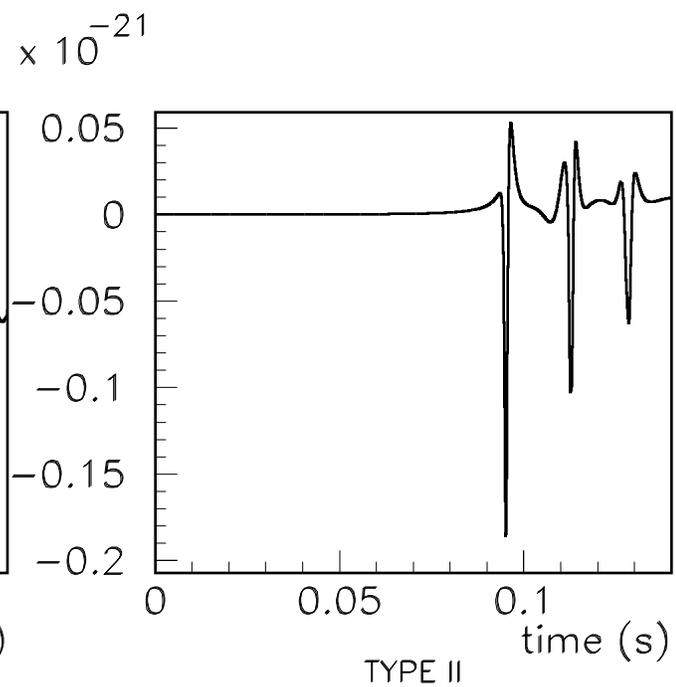
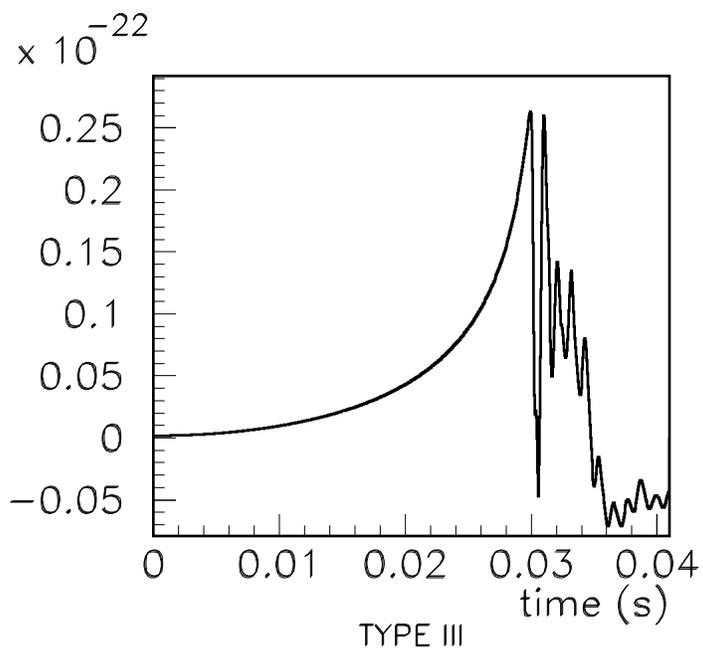
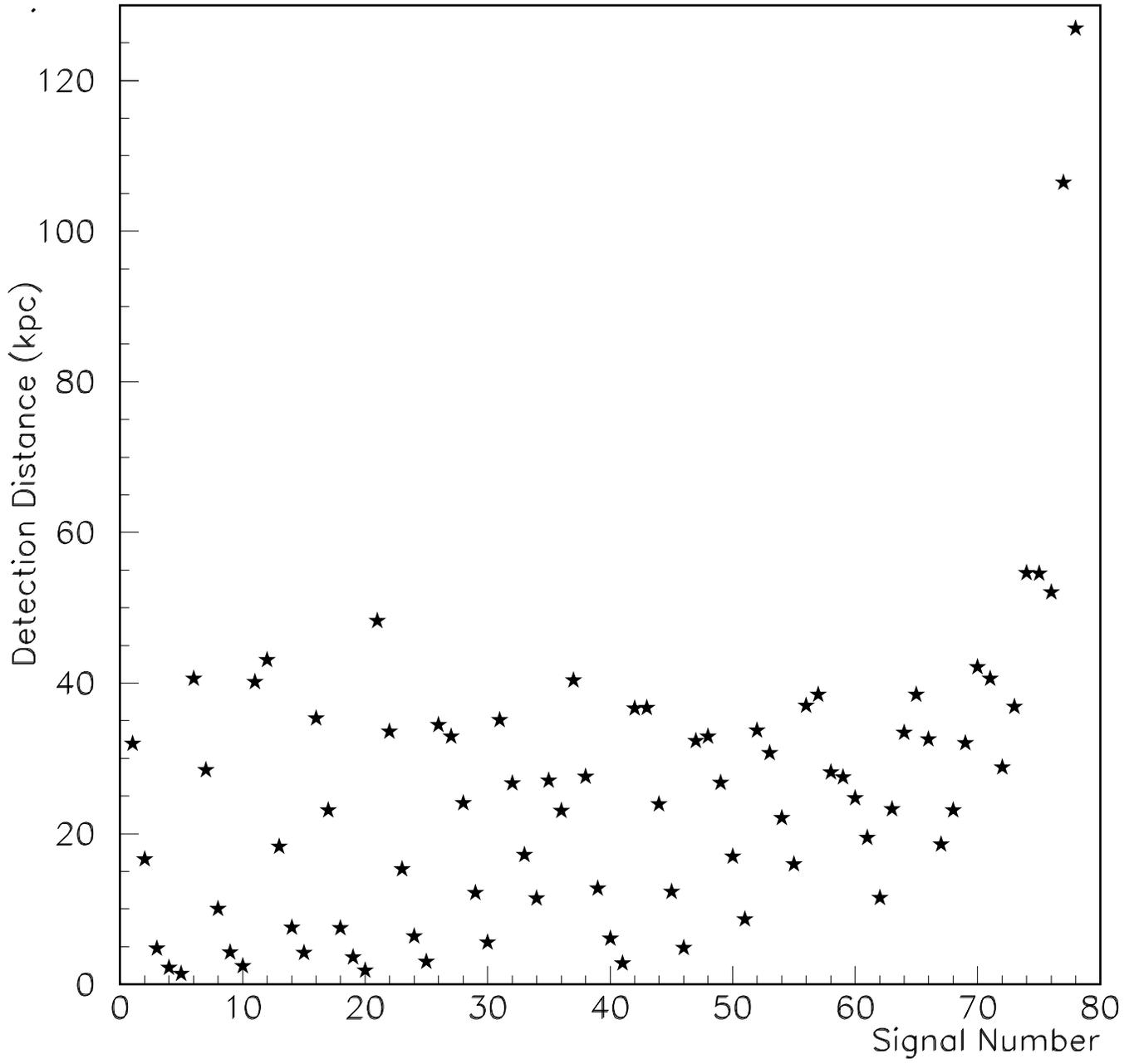
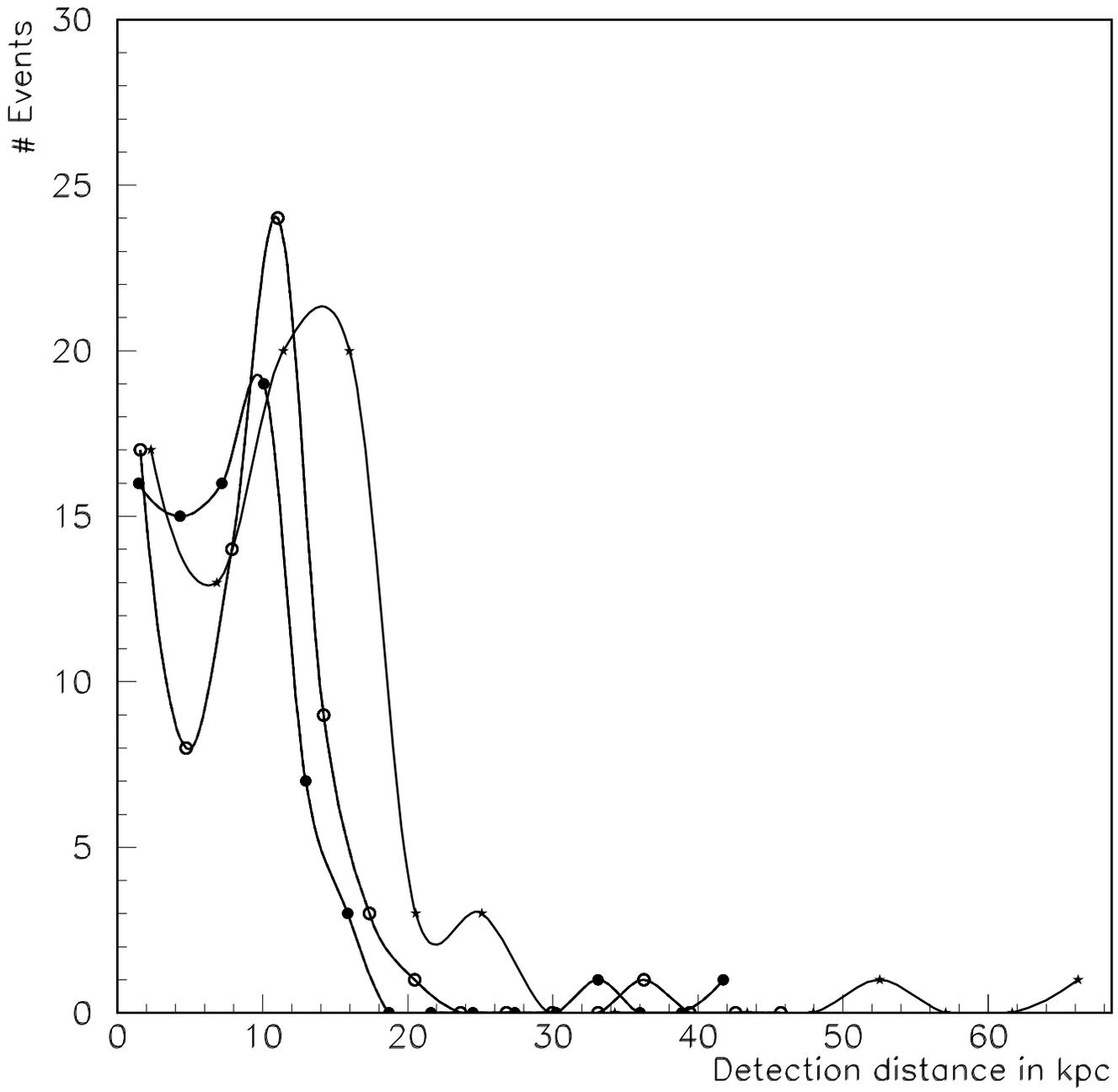
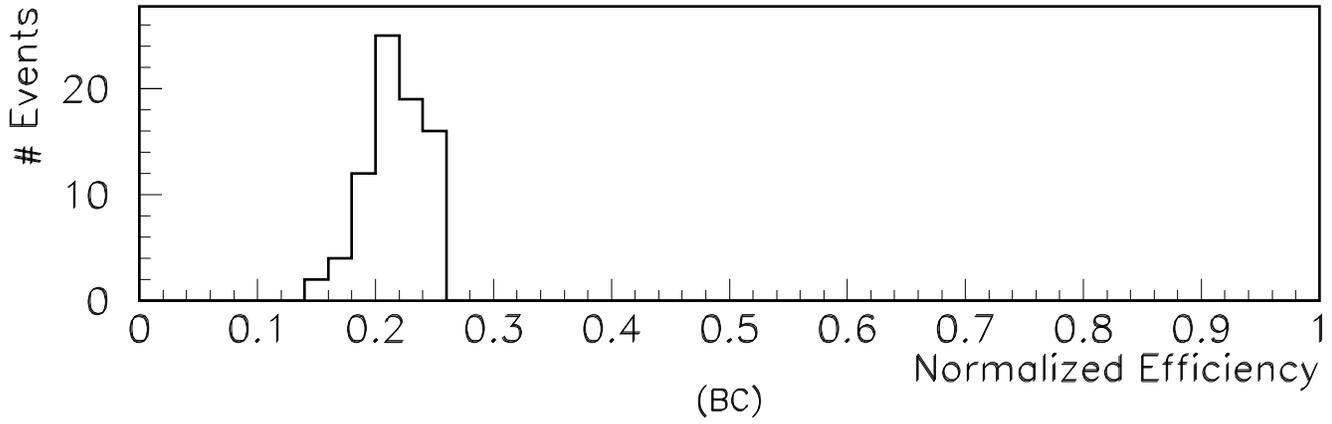
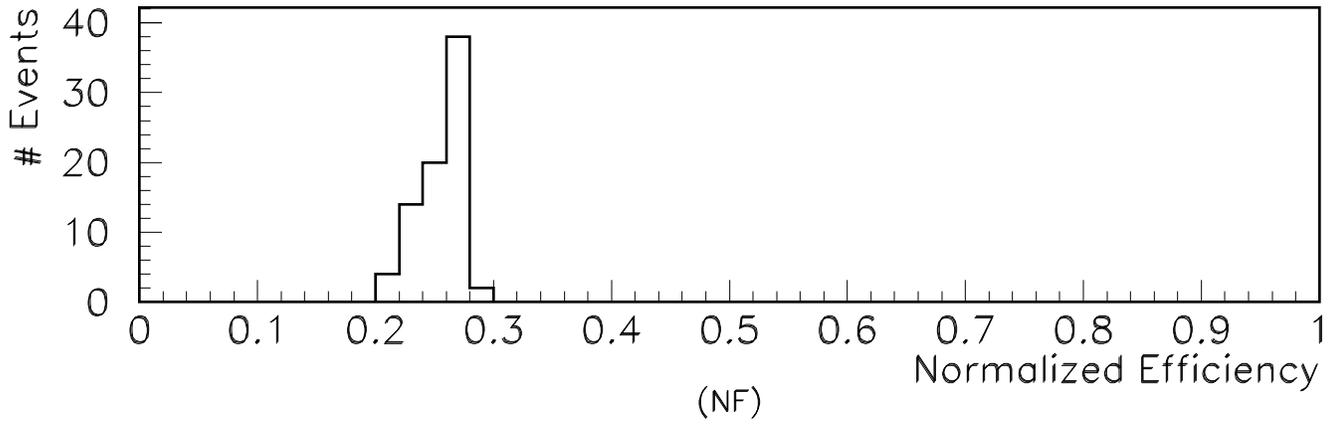
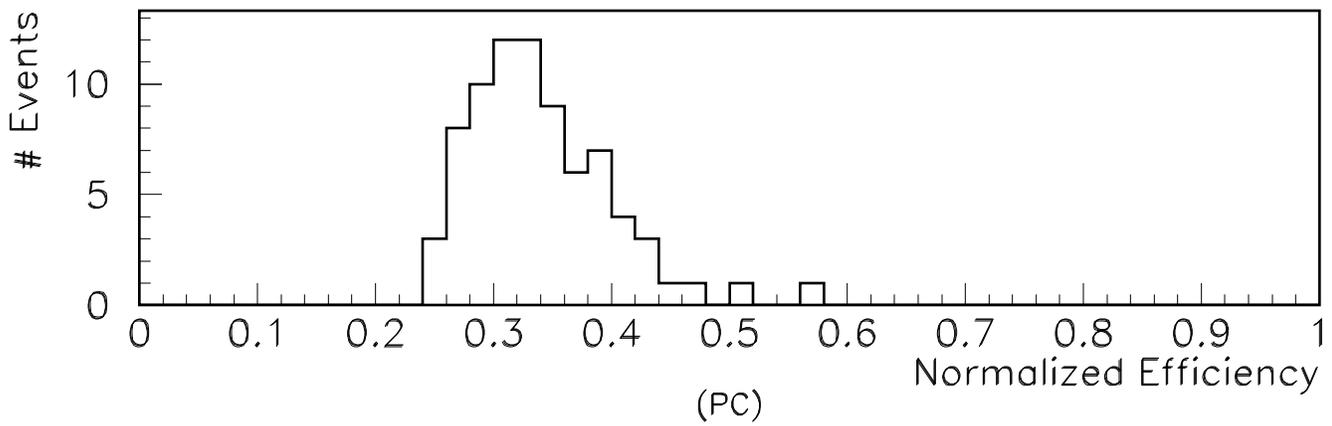
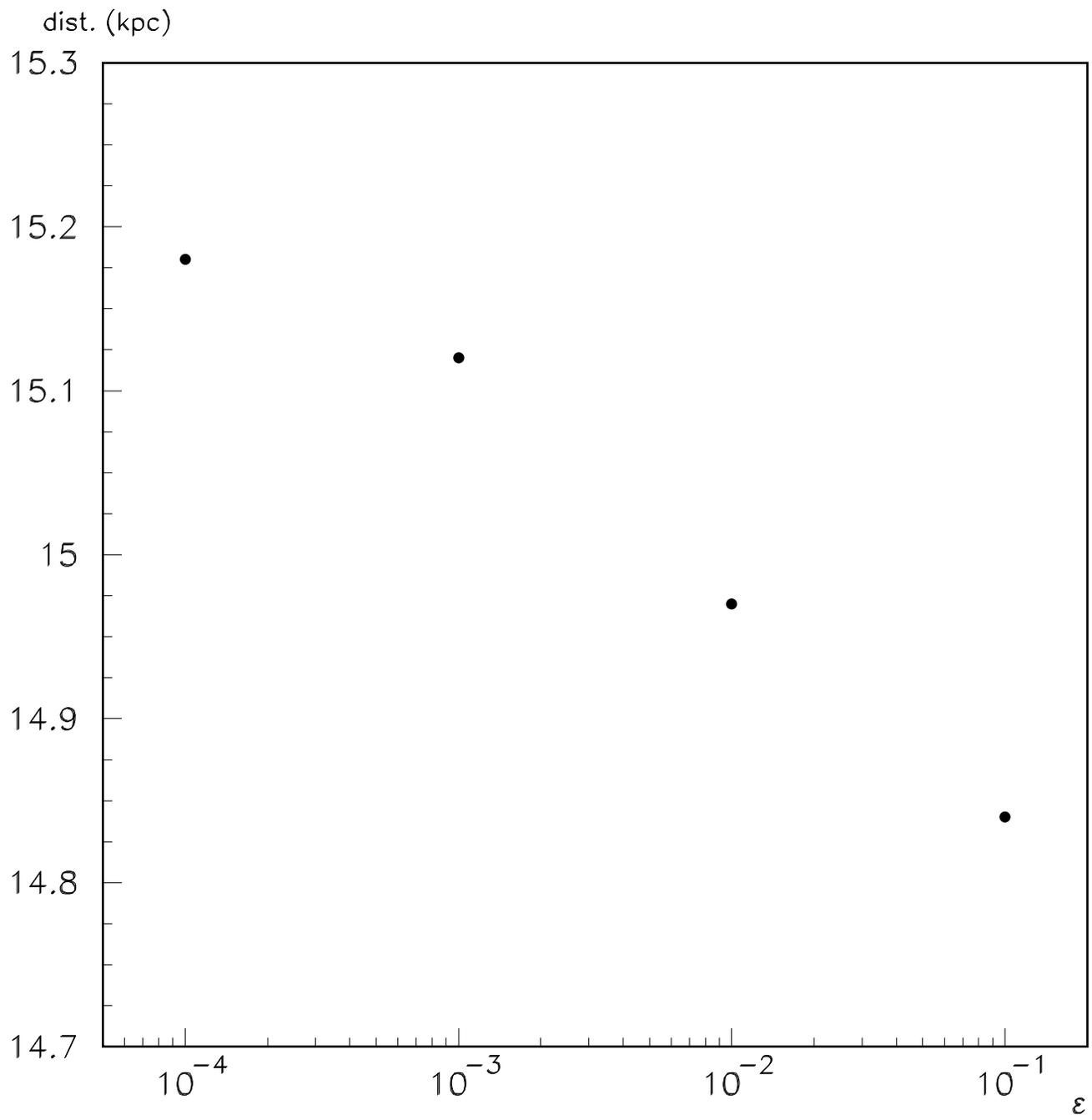
\end{document}